\documentclass[11pt]{article}
\usepackage[pdftex,colorlinks=true,urlcolor=black,filecolor=black,linkcolor=black,
            pdftitle={Halo Spacetime.},
            pdfauthor={Mark D. Roberts},
            pdfsubject={relativity, galaxies},
            pdfkeywords={galaxies},
            pagebackref,pdfpagemode=None,bookmarksopen=true]{hyperref}
\usepackage{amssymb}
\usepackage{graphicx}
\usepackage{epsfig}
\usepackage{color}
\newcommand{\bc}{\begin{center}}
\newcommand{\ec}{\end{center}}
\newcommand{\bi}{\begin{itemize}}     
\newcommand{\ei}{\end{itemize}}
\newcommand{\bd}{\begin{description}} 
\newcommand{\ed}{\end{description}}
\newcommand{\bn}{\begin{enumerate}}   
\newcommand{\en}{\end{enumerate}}
\newcommand{\be}{\begin{equation}}
\newcommand{\ee}{\end{equation}}
\newcommand{\ber}{\begin{eqnarray}}
\newcommand{\ear}{\end{eqnarray}}
\newcommand{\ba}{\begin{array}}
\newcommand{\ea}{\end{array}}
\newcommand{\bv}{\begin{verbatim}}
\newcommand{\ev}{\end{verbatim}}

\newcommand{\bx}{\Box}

\newcommand{\fr}{\frac}

\newcommand{\Lg}{{\cal L}}
\newcommand{\n}{\nonumber\\}

\newcommand{\p}{\partial}
\begin{document}
\title{Halo Spacetime.}
\author{Mark D. Roberts,\\
\href{http://www.violinist.com/directory/bio.cfm?member=robemark}
{54 Grantley Avenue,  Wonersh Park,  GU5 0QN,  UK},\\
robemark@gmail.com
}
\date{$30^{th}$ of January 2011}
\maketitle
\begin{abstract}
It is shown that constant galactic rotation curves require a logarithmic potential
in both Newtonian and relativistic theory.
In Newtonian theory the density vanishes asymptotically,
but there are a variety of possibilities for perfect fluid Einstein theory.
\end{abstract}
{\tiny\tableofcontents}
\section{Introduction.}\label{intro}
Constant galactic rotation curves require a logarithmic potential in both newtonian
and relativistic theory.
The logarithmic potential contrasts with
heuristic derivations of the Tully-Fischer \cite{carmeli,hartnett} relationship,
which require $\phi\approx1/r$.
As pointed out in \cite{mdr16} for a potential $\approx\ln(r)$
one would expect as well a metric with $\approx\ln(r)$ terms
and hence first derivatives and Christoffel symbol corresponding to the metric
are of order $\approx1/r$;
thus both the derivatives and products of the Christoffel symbol are of order $\approx1/r^2$
so that linear approximations to the Riemann tensor
and hence the field equations are not necessarily consistent.
This is why for a relativistic analysis one has to start again
with the properties of geodesics,
this is done here in section \ref{geodesics}.
A problem with any model of galactic rotation is what happens asymptotically.
The logarithmic potential necessary for constant rotation curves
diverges as $r\rightarrow\infty$.
One way around a divergent potential
seems to be to invoke a ``halo'' which is a region surrounding
a galaxy where there is an approximately logarithmic newtonian potential.
The properties of the ``halo'' have some sort of unspecified cut off
so that the potential does go out to too large distances,
quite how this works is not at all clear,
after all binary and multiple galaxies exhibit unusual dynamics,
as do stellar clusters,
but presumably nearby galaxies have more gravitational influence than distant ones,
so that at some distance a logarithmic potential must have a cut off.
Another way of looking at what happens asymptotically
for a logarithmic potential is to note that it is
only the derivatives of the potential that have effects and $\phi_r\approx1/r$
does not diverge asymptotically.
Yet another way is to introduce smooth step functions, see section \ref{asymptotics}.
``Halo'' is a term that is usually restricted to an area of ``dark matter'' surrounding
a galaxy;  however if a model uses matter fields or different field equations to model
galactic rotation one would again require a region that these fields model and it is
simpler to again refer to this as a ``halo'' rather than use new terminology.

In section \ref{fieldequations} the analysis is generalized to the perfect fluid
Einstein equations where it is shown that
the density is the same as in the Newtonian analysis,
but with additional terms appearing in the pressure and the density of order $v_c^4$.
The analysis requires four assumptions:
1)a galactic halo can be modelled by a static spherically symmetric spacetime,
2)in the limit of no rotation halo spacetime is Minkowski spacetime,
3)the constraint on the first derivatives of the metric needed in order to produce
constant rotation curves can be applied,
4)the perfect fluid Einstein equations;
the density is the same as in the Newtonian analysis,
but with additional terms appearing in the pressure and the density of order $v_c^4$.
The assumption 1) is that a galactic halo
can be modelled by a static spherically symmetric spacetime.
That it is static is a good approximation as dynamical disturbances in galaxies
do not seem to be the cause of constant rotation curves.
That it is spherically symmetric seems to be born out by observations \cite{merrifield}.
Thus it is taken that a spherically symmetric halo causes constant galactic rotation,
rather than the rotation of the luminous part of the galaxy.
Using spherical symmetry as opposed to axial symmetry considerably simplifies calculations.
There are {\it two} problems with the assumption 2) that in the limit of no rotation
the spacetime is Minkowski spacetime.
The {\it first} is that it excludes the possibility of other limiting structure,
such as deSitter spacetime.
The {\it second} is that even non-rotating galaxies have non-flat spacetimes:
in other words this assumption is that to lowest order the rotating part of the
description of a galaxies spacetime can be described independently of its
non-rotating part.
The assumption 3) is the constant velocity rotation constraint,
this is a constraint on the first derivatives of the metric,
it has been derived in \cite{mdr37},  and is derived again in the next section.
For spherical symmetry it is an unusual constraint because it often involves
logarithms rather than powers of the metric.
The assumption 4) is the perfect fluid Einstein equations.
For spherical symmetry the stress has three arbitrary stress functions $(\rho,p_r,p_\theta)$.
The assumption that the stress is a perfect fluid reduces the three components
to two $(\rho,p)$ via the isotropic pressure equation $p=p_r=p_\theta$.
For a perfect fluid the density $\rho$ and the pressure $p$
are independent unless an equation of state is imposed.
The reduction of the three arbitrary stress functions to two can be thought of
as a second constraint on a first derivative of the metric.
The fourth assumption has two aspects:
firstly that Einstein's equations are the correct ones to model galaxies,
and secondly that any ''dark matter'' has isotropic pressure so that $p_r=p_\theta$.
In section \ref{fieldequations} whether it is permissible to expand in the
velocity of rotation $v_c/c$, where $c$ is the velocity of light,
rather than a radial parameter $r$ is also discussed.
For a galaxy it is not clear what radial parameter could be used because of the
problem of what happens as $r\rightarrow\infty$,
for the luminosity radial parameter $R$ one has that the metric $\approx \ln(R)$
suggesting that it is not a good parameter to expand in.
In both cases the expansions are independent of the mass of the galaxy $m$
which does not enter explicitly into calculations here to lowest order.
Taking galactic rotation to be in any way related to the Kerr solution is always
misleading, as Kerr rotation is a short range effect and galactic rotation is a long
range effect.

A selection of recent literature on rotation curves can be divided into three types.
The first are papers which use the MOND modification of gravity,
see for example \cite{bekenstein,BB}.
The second are papers which invoke higher dimensions and/or branes,
see for example \cite{HC,HK,KS,MH,PBK}.
The third are papers which invoke new fields,  or field equations, or potentials,
see for example \cite{mdr2,mdr6,mdr16,echardt,kislev,ALS,CT,VL,HMSV,ZN}.

Conventions include:  metric -+++ and field equations $G_{ab}=8\pi GT_{ab}$
Calculations were done using maple9.
\section{Newtonian theory.}\label{newt}
First a newtonian analysis is given,
the presentation follows \cite{mdr37},  but is shorter.
In spherical coordinates newton's second law has two dynamical components
\be
-\left(\phi_r,\phi_\theta\right)=
\left(\ddot{r}-r\dot{\theta}^2,\fr{d}{dr}(r^2\dot{\theta})\right),
\label{nl2}
\ee
where $\phi$ is the newtonian potential and the subscript indicates which coordinate it
is differentiated with.
The $\theta$ component integrates to give
\be
r^2\dot{\theta}=L,
\label{tcomp}
\ee
where $L$ is the angular momentum for each individual particle.
The square of the velocity at any point is
\be
v^2=\dot{r}^2+(r\dot{\theta})^2.
\label{vel}
\ee
For circular orbits $\dot{r}=0$ so that (\ref{vel}) becomes
\be
v_c^2=(r\dot{\theta})^2=\fr{L^2}{r^2},
\label{vc}
\ee
also for circular orbits newtons second law (\ref{nl2}) gives
\be
\phi_r=r\dot{\theta}^2=\fr{L^2}{r^3}.
\label{circn2}
\ee
Eliminating $L^2$ from these two equations (\ref{vc}) and (\ref{circn2}) gives
\be
\phi_r=\fr{v_c^2}{r},
\label{noL}
\ee
integrating gives the newtonian potential
\be
\phi=v_c^2\ln(r),
\label{npot}
\ee
where $v_c$ is a constant.
Poisson's equation is
\be
4\pi G\rho=\bx\phi=\fr{1}{r^2}\left(r^2\phi_r\right)_r=\fr{v_c^2}{r^2}.
\label{poissoneq}
\ee
\section{The geodesics of rotation curves.}\label{geodesics}
The constant rotation curve constraint can be derived as follows.
The axi-symmetric line element can be taken as
\be
ds^2=g_{tt}dt^2+2g_{t\phi}dtd\phi+g_{\phi\phi}d\phi^2+g_{rr}dr^2+g_{\theta\theta}d\theta^2.
\label{aximetric}
\ee
The geodesic lagrangian is
\be
2\Lg=g_{tt}\dot{t}^2+2g_{t\phi}\dot{t}\dot{\phi}+g_{\phi\phi}\dot{\phi}^2
+g_{rr}\dot{r}^2+g_{\theta\theta}\dot{\theta}^2=p_ap^a,
\label{geolag}
\ee
where $2\Lg=-1$ for timelike geodesics.
The euler equations are
\be
\fr{d p_a}{d \tau}=\fr{\p\Lg}{\p x^a}.
\label{euler}
\ee
For (\ref{geolag})
\be
p_r\equiv\fr{\p\Lg}{\p\dot{r}}=g_{rr}\dot{r},~~~
2\fr{\p\Lg}{\p r}=g_{tt}'\dot{t}^2+2g_{t\phi}'\dot{t}\dot{\phi}
+g_{\phi\phi}'\dot{\phi}^2+g_{rr}'\dot{r}^2+g_{\theta\theta}'\dot{\theta}^2,
\label{rpart}
\ee
where $f'$ denotes $\p f/\p r$.
Thus the $r$ component of the euler equation is
\be
\fr{dp_r}{d\tau}=g_{rr}\ddot{r}+g_{rr}'\dot{r}^2+g_{rr,\theta}\dot{r}\dot{\theta}=
g_{tt}'\dot{t}^2+2g_{t\phi}'\dot{t}\dot{\phi}+g_{\phi\phi}'\dot{\phi}^2
+g_{rr}'\dot{r}^2+g_{\theta\theta}'\dot{\theta}^2,
\label{reuler}
\ee
with a similar equation for the $\theta$ component.
If required the $p_t$ and $p_\phi$ components can be expressed in terms of the energy $E$ and
momentum $L$ for each geodesic,  and then with \ref{reuler} this allows generalizations of the
Binet equation;  however for present purposes this is unnecessary as
the problem much simplifies because constant velocity curves require
\be
\dot{\theta}=\dot{r}=0.
\label{consrot}
\ee
Substituting into the lagrangain (\ref{geolag})
and the $r$ \& $\theta$ components of the euler equation (\ref{reuler}) gives
\ber
&2\Lg=g_{tt}\dot{t}^2+2g_{t\phi}\dot{t}\dot{\phi}+g_{\phi\phi}\dot{\phi}^2,
\label{redlag}
\\
&0=g_{tt}'\dot{t}^2+2g_{t\phi}'\dot{t}\dot{\phi}+g_{\phi\phi}'\dot{\phi}^2,
\nonumber
\ear
respectively.
There is a similar equation to the second of (\ref{redlag}) with $f'$ denoting $\p f/\p\theta$.
The angular momentum is
\be
\Omega\equiv\fr{\dot{\phi}}{\dot{t}}=\fr{v_c}{\sqrt{g_{\phi\phi}}},
\label{eq18}
\ee
note that this corrects equation (27) of \cite{mdr37}
where there is an $r$ rather than a $\sqrt{g_{\phi\phi}}$
in the denominator of the last term.
Substituting into the lagrangian equation of (\ref{redlag})
from which $\dot{t}$ can be calculated,
this is not necessary for present purposes.
Substituting into (\ref{redlag}) first for the first equality in (\ref{eq18})
and then for the second equality in (\ref{eq18}) gives
\be
0=g_{tt}'+2g_{t\phi}'\Omega+g_{\phi\phi}'\Omega^2
=g_{tt}'+2\fr{v_c}{\sqrt{g_{\phi\phi}}}g_{t\phi}+\fr{v_c^2g_{\phi\phi}'}{g_{\phi\phi}}.
\label{constgeos}
\ee
There is the possibility the galactic rotation curves are intrinsically axi-symmetric and cannot be
modelled by spherical symmetry,  however the newtonian model with potential \ref{npot} is
spherically symmetric and if there is a requirement that this is a limit of a relativistic model
then to lowest order this must also be spherically symmetric.
For spherical symmetry with line element
\be
ds^2=-\exp(2\nu)dt^2+\exp(2\mu)dr^2+\exp(2\psi)d\Sigma^2_2,
\label{ssst}
\ee
where $d\Sigma^2_2=d\theta^2+\sin(\theta)^2d\phi^2$,
the constraint (\ref{constgeos}) reduces to
\be
\left\{\exp(2\nu)\right\}'=2v_c^2\psi'.
\label{ssconstraint}
\ee
The requirement that for $v_c=0$ the metric is Minkowski fixes the constant of integration;
integrating the metric becomes
\be
ds^2=-\left(1+2v_c^2\psi(r)\right)dt^2+\exp(2\mu(r))dr^2+\exp(2\psi(r))d\Sigma_2^2.
\label{ss2fn}
\ee
At first sight this line element has two arbitrary function in it,
however defining the luminosity distance $r_{lm}$
\be
r_{lm}\equiv\exp(\psi(r)),~~~
\exp(\mu_{lm}(r_{lm})\equiv\fr{\exp(\mu({\rm inv}\psi(\ln(r_{lm}))))}{r_{lm}^2\psi_r^2},
\label{redefr}
\ee
and dropping the subscript $lm$ the line element becomes
\be
ds^2=-(1+2v_c^2\ln(r))dt^2+\exp(2\mu)dr^2+r^2d\Sigma^2_2,
\label{ss1fn}
\ee
leaving just one arbitrary function $\mu(r)$.
This can be fixed by using field equations.
\section{The perfect fluid Einstein equations.}
\label{fieldequations}
For a spherically symmetric spacetime of the form (\ref{ss1fn}) there are three non-vanishing
components of the Einstein tensor
\ber
8\pi G\rho&=&-G^t_{.t}=
\fr{1}{r^2}(1-(r\exp(-2\mu))'),\label{et}\\
8\pi Gp_r&=&+G^r_{.r}=
\fr{1}{r^2}\left(-1+\exp(-2\mu)\fr{(1+2v_c^2+2v_c^2\ln(r))}{(1+2v_c^2\ln(r))}\right)\n
8\pi Gp_\theta&=&+G^\theta_{.\theta}=
\fr{-\exp(-2\mu)}{r^2(1+2v_c^2\ln(r))}
\left(\fr{v_c^4}{(1+2v_c^2\ln(r))}+r\mu'(1+v_c^2+2v_c^2\ln(r))\right)
\nonumber
\ear
where $G^\phi_{.\phi}=G^\theta_{.\theta}$,  $\rho$ is the density,
$p_r$ is the radial pressure,  and $p_\theta$ is the angular pressure.
These field equations are simple as they linear in only the first derivatives of $\mu$
and so should have many solution,
for present purposes just the case of a perfect fluid is examined.
The requirement that the stress be that of a perfect fluid is that
the pressure is isotropic $p=p_r=p_\theta$,
and this is sufficient to completely determine the metric.
Subtracting $p_\theta$ from $p_r$ gives a first order differential equation in $\mu$,
\be
r\mu'+\fr{1+v_c^2+2v_c^2\ln(r)}{1+2v_c^2\ln(r)}
+\exp(2\mu)\left(r^2(p_\theta-p_r)-1\right)\fr{1+2v_c^2\ln(r)}{1+v_c^2+2v_c^2\ln(r)}=0,
\label{ode}
\ee
setting $p_\theta=p_r$ and solving gives
\be
ds^2=-(1-2v_c^2\ln(r))dt^2+\fr{dr^2}{2r^2(k-I)(1+2v_c^2\ln(r))}+r^2d\Sigma^2_2,
\label{newle}
\ee
the requirement that the line element reduces to Minkowski
when the velocity vanishes gives $k=0$.
$I$ is the integral
\ber
I&\equiv&\int\fr{dr}{r^3}\fr{1}{(v_c^2+1+2v_c^2\ln(r))}\\
&=&-\fr{1}{2r^2}+\left(1+\ln(r)\right)\fr{v_c^2}{r^2}
-\left(5+8\ln(r)+4\ln(r)^2\right)\fr{v_c^4}{2r^2}+{\cal O}\left(v_c^6\right).
\nonumber
\label{integral}
\ear
For $k=0$ the pressure is
\be
8\pi Gp=-\fr{1}{r^2}+2\left(1+2v_c^2+2v_c^2\ln(r)\right)(k-I)=\fr{v_c^4}{r^2}
+{\cal O}\left(v_c^6\right),
\label{pressure}
\ee
and the density is
\ber
\label{density}
8\pi G\rho&=&\fr{(3+v_c^2+6v_c^2\ln(r))}{(1+v_c^2+2v_c^2\ln(r))r^2}
+2\left(3+2v_c^2+6v_c^2\ln(r)\right)(I-k)\\
&=&+\fr{2v_c^2}{r^2}-\left(9+4\ln(r)\right)\fr{v_c^4}{r^2}+{\cal O}\left(v_c^6\right).
\nonumber
\ear
The ${\cal O}(v_c^0)$ term of the integral (\ref{integral}) shows that the spacetime
is Minkowski when $v_c=k=0$,  for $k\ne0$ the $g_{rr}$ component looks similar to deSitter
spacetime however there is no corresponding term in $g_{tt}$
and any such term would violate the the rotation constraint (\ref{constgeos}),
here usually $k=0$.
The expansion in $v_c$ shows that to lowest order the density (\ref{density})
agrees with the newtonian expression (\ref{poissoneq}),
however the expansion does not always converge,
terms becoming larger and changing sign for each increase in $v_c^2$.
The large distance $r$ properties of \ref{density} cannot be obtained from the $v_c^2$
expansion as this is not the same as a $r^{-2}$ expansion because of the $\ln(r)$ terms.
The technical reason that a $r^{-2}$ expansion cannot be obtained is that there is no series
expansion for the logarithmic integral $J\equiv\int dx/\ln(x)$,  if it is attempted each term
is of the same order.  This makes it necessary to work with the exponential integral defined by
\be
Ei(a,z)\equiv\int^\infty_1 \frac{d\ell}{\ell^a}\exp\left(-\ell z\right),
\label{eidef}
\ee
and illustrated in the figure.
\begin{figure}
\includegraphics[height=12cm]{figliei.jpg}
\caption{The logarithmic and exponential integrals.}
\end{figure}

Using this definition \ref{eidef} the integral \ref{integral} can be expressed as
\be
I=-\fr{1}{2v_c^2}\exp\left(1+\fr{1}{v_c^2}\right){\rm Ei}\left(1,1+\fr{1}{v_c^2}+2\ln(r)\right),
\label{ei}
\ee
and this can be used to numerically produce graphs of the properties of the spacetime.
\begin{figure}
\includegraphics[height=12cm]{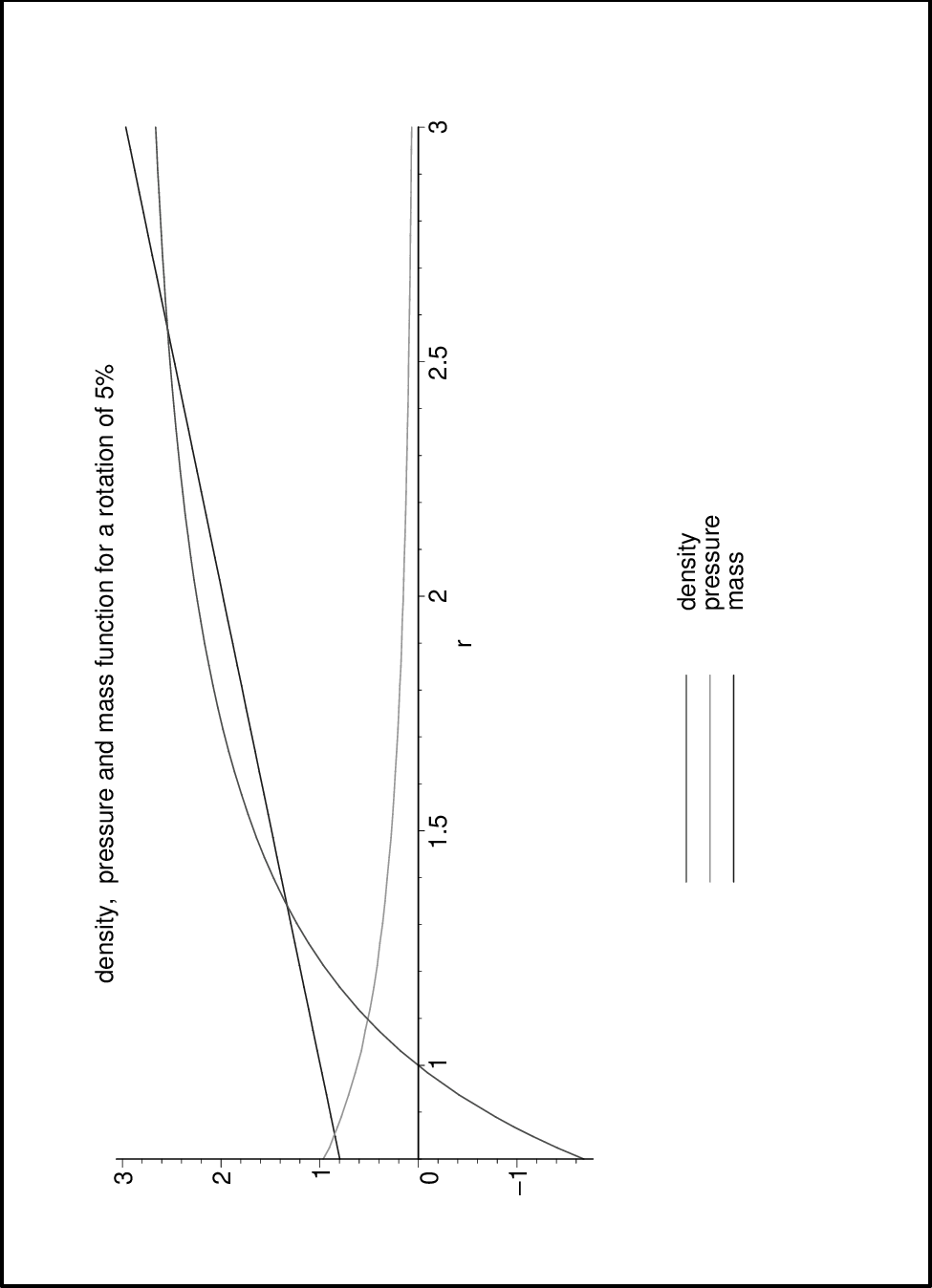}
\caption{The density,  pressure and mass function.}
\end{figure}
In figure one the density,  $10^5\times$pressure and $400\times$mass function
are plotted for $v_c/c=0.05$ (not $v_c^2=0.05$) and $k=0$.
Asymptotically the density approaches $3$,  the pressure $0$,  and the mass function $r$.
In the region $v_c=0.01-0.02$ the machine overflows,
which is unfortunate as $v_c=10^{-6}$ is a more realistic value.
For $k=1$ the density and mass function appear negative,
the critical value for which positive density is restored is around $k=2.5\times10^{-3}$,
there does not seem to be a critical value for the mass function which always goes
negative for large enough $r$.
For $k=-1$ the pressure is negative,
the critical density for which positive pressure is restored is around $k=-2.5\times10^{-10}$.

\begin{figure}
\includegraphics[height=12cm]{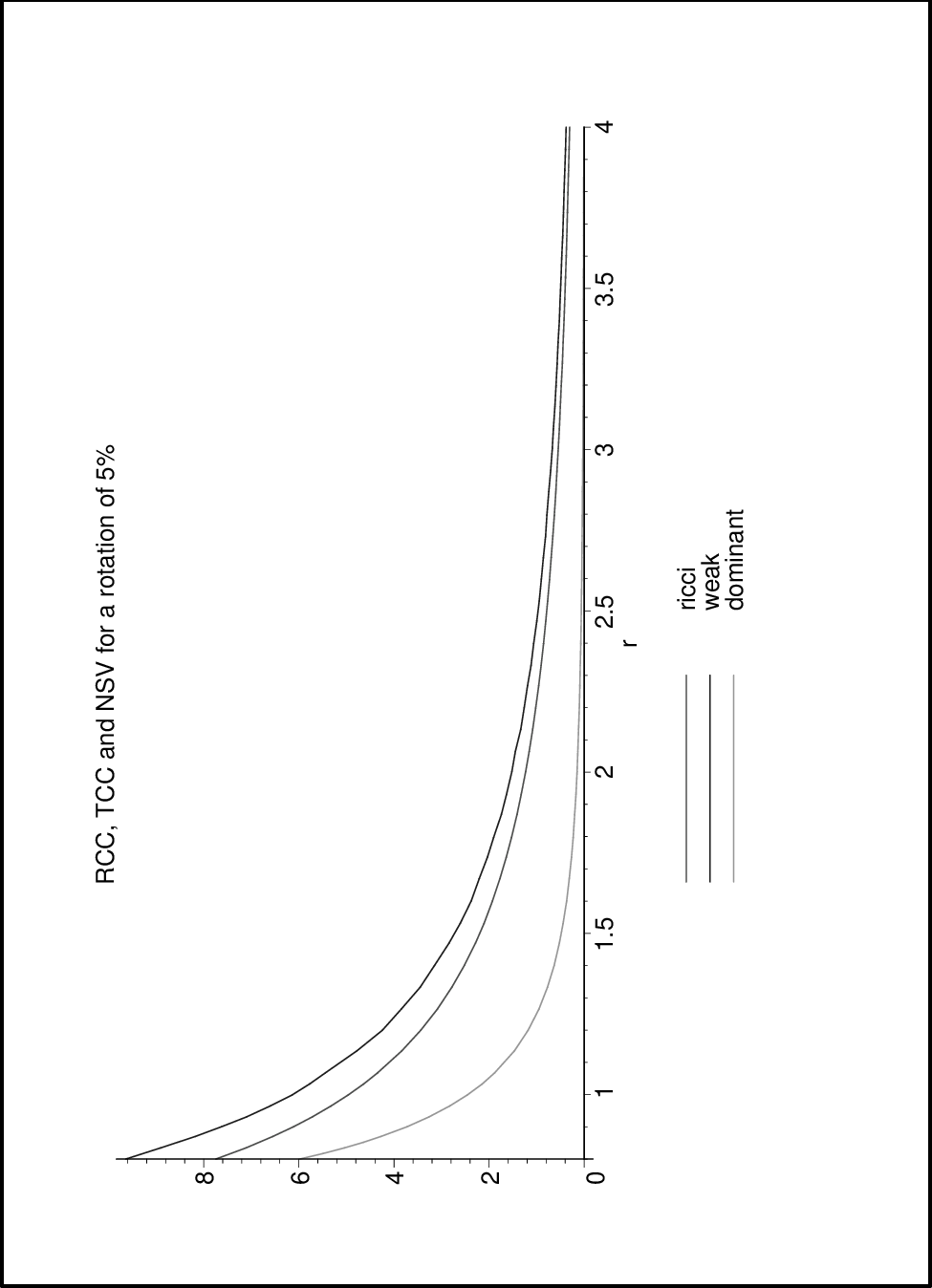}
\caption{The energy conditions.}
\end{figure}
The null convergence condition is $RCC\equiv R_{ab}n^an^b\ge0$,
for $n_a$ a null vector p.95\cite{HE},
the weak energy condition is $TCC\equiv T_{ab}V^aV^b$,
for $V_a$ a timelike vector p.89\cite{HE},
the dominant energy condition is the timelike convergence condition and
$T^{ab}V_a$ is a non-spacelike vector p91\cite{HE},
NSV is the size of $T^{ab}V_a$.
In figure two $10^3\times$RCC and $10^6\times$TCC and $-10^5\times$NSV
are plotted for $v_c=0.05$ and $k=0$.
For $k=1$ the null convergence condition is violated,
the critical value around which it seems to be restored is $k=10^{-6}$.
For $k=-1$ the weak energy condition is violated and this seems always to be the case for
large enough $r$.
\begin{figure}
\includegraphics[height=12cm]{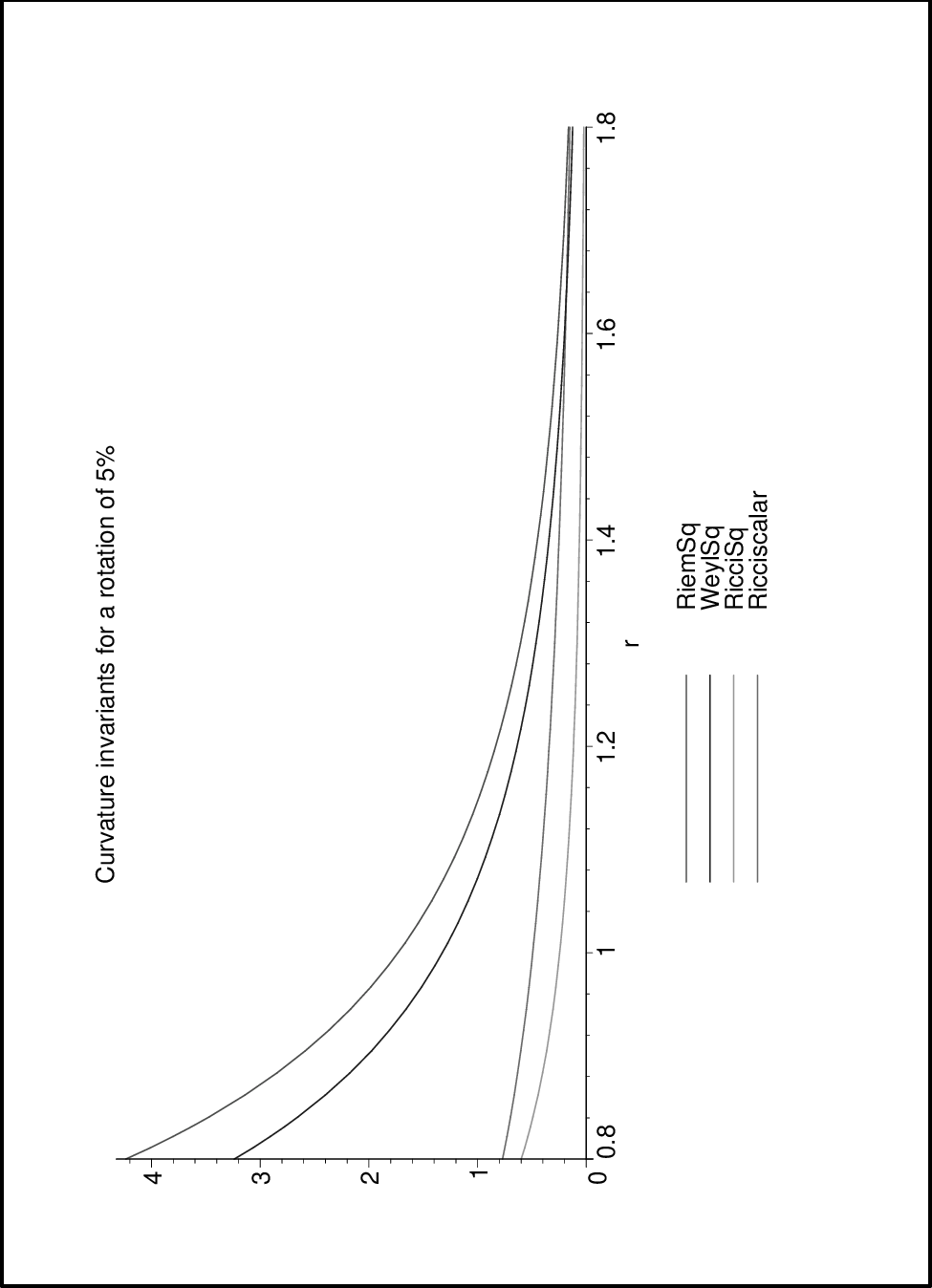}
\caption{The curvature invariants.}
\end{figure}
The curvature invariant RiemSq is defined by $RiemSq=R_{abcd}R^{abcd}$,
with similar definitions for WeylSq and RicciSq.
In figure three $10^4\times$RiemSq, $10^4\times$WeylSq, $10^4\times$RicciSq
and $10^2\times$Ricciscalar (not squared) are plotted for $v_c=0.05$ and $k=0$.
It does not seem to be possible numerically to determine whether the divergence happens at
$r=\exp(-1/(2v_c^2))\approx 10^{-86}$ or $r=0$ or both.
A surprising feature of figure three is that WeylSq is large compare to RicciSq,
this has the interpretation, see \cite{HE}p.85,
that more of the curvature is due to gravity as opposed to matter.
For $k=\pm1$ the RicciSq is large compared to WeylSq,
the critical value seems to be around $k=\pm10^{-4}$,
but seems to highly dependent on $r$.
\newpage
\section{Asymptotics and units.}\label{asymptotics}
Constant velocity curves are only observed over a certain region,
how short and long radial distances fit to this region is a problem.
There seem to be three approaches to this:
the {\it first} is to adjust things at the last moment and produce an onion model in which
the constant velocity region as given has other spacetimes are fixed to it,  see \cite{mdr39},
the {\it second} is to adjust things in the middle by choosing contrived newtonian potentials,
the {\it third} is to adjust things at the beginning by letting $v_c\rightarrow v_c(r)$.

Looking at the third approach first,
for a newtonian model,  equations (\ref{noL}) and (\ref{poissoneq}) become
\be
r\phi_r=v_c^2(r),~~~4\pi G\rho=\fr{1}{r^2}\left(rv_c^2(r)\right)'.
\label{np}
\ee
\begin{figure}
\includegraphics[height=8cm]{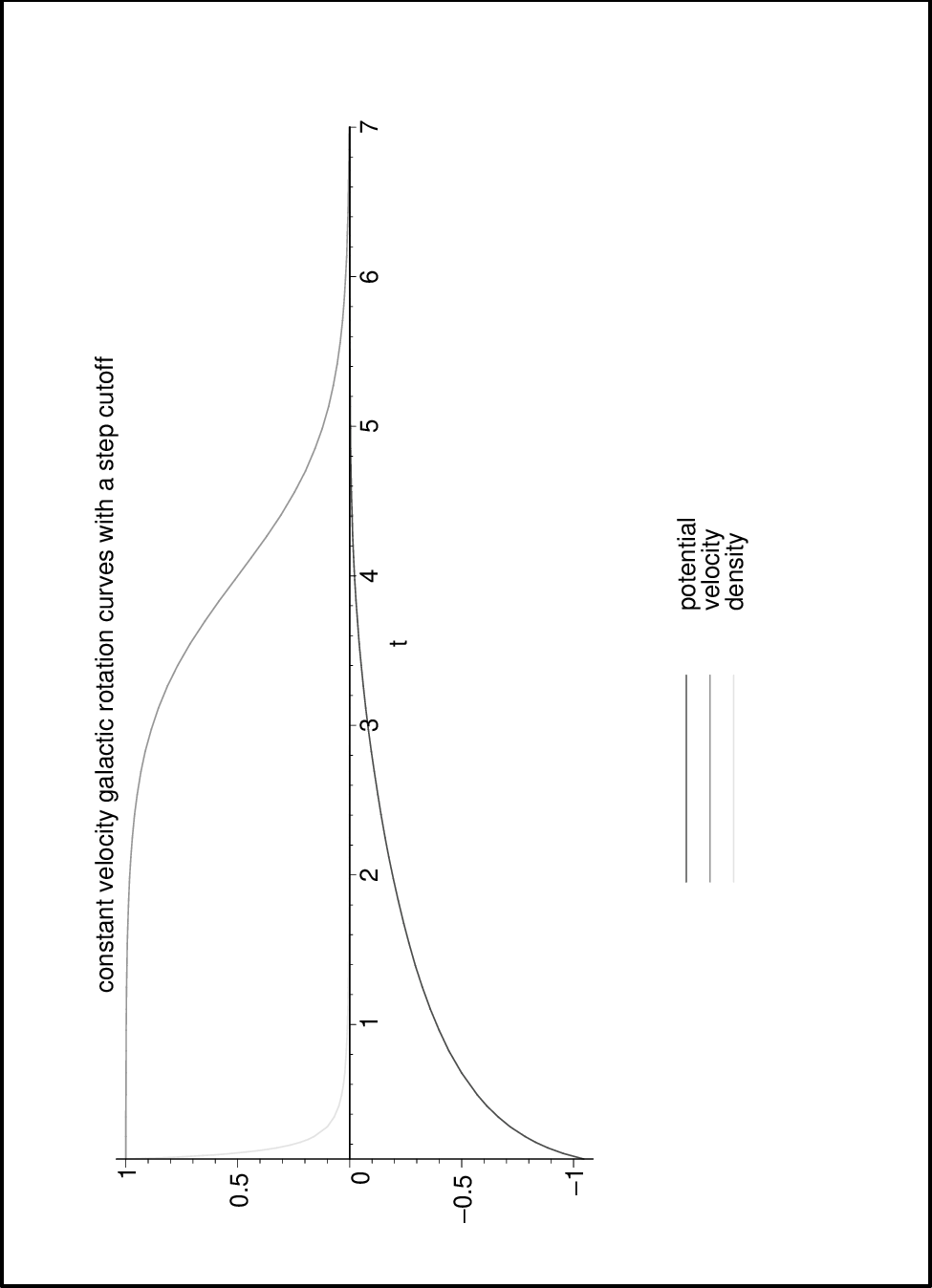}
\caption{A good newtonian potential.}
\end{figure}
For purposes of illustration consider the smoothed out step around $r=r_o$
\be
S(r,r_o)\equiv\fr{1}{2}\left(1-\tanh(r-r_o)\right)=\fr{1}{\exp(2(r-r_0))+1}.
\label{step}
\ee
Apply the step (\ref{step}) to the velocity
\be
v_c^2(r,r_o)=S(r,r_o),~~~
\label{goodvel}
\ee
then the potential is given by an integral, (\ref{gvp}) with $k=1$,
The integral (\ref{gvp}) has to be evaluated numerically
\be
\phi=\int\fr{dr}{r(\exp(2(r-r_o)+k)},
\label{gvp}
\ee
for the velocity (\ref{goodvel}),  (\ref{np}) gives the density
\be
4\pi G\rho=\fr{1+(1-2r)\exp(2(r-r_o)}{r^2(1+\exp(2(r-r_o)))^2}
\label{gdensity}
\ee
For $r_o=4$ the velocity (\ref{goodvel}),
the potential (\ref{gvp}) divided by $3.5$
and the density (\ref{gdensity}) divided by $100$
are plotted in figure four,
to avoid singularities at the origin the integral is taken from $0.1$ rather than $0$.
By comparison with the integral
$\int dr/(\exp(2(r-r_o)+1)=\ln(\exp(2(r-ro))/(\exp(2(r-ro))+1))/2$
the potential converges to $0$ as $r\rightarrow\infty$,  which is good for two reasons,
firstly that it converges at all unlike $\ln(r)$,
secondly that the limit is $0$ so that there is no trace of the potential
at great distance from the galaxy;
similarly comparing with the integral $\int dr/r=\ln(r)$
the potential is seen to diverge for short distances.
Surprisingly both the potential and the density show no unusual properties at the step $r_o$.

For the second approach apply the step \ref{step} to the potential so that
\be
\phi=\ln(r)S(r),~~~
v_c^2:=r(\ln(r)S(r))'
\label{goodpot}
\ee
the potential starts out like a $\ln$ potential and then smooths out,
however $v_c^2$ becomes negative which is unphysical.

For short distance behaviour one can again use smoothed out step functions,
however numerical studies show that there are rings of large density near the
smoothed out step.
To produce a relativistic model with potential (\ref{gvp}) is a substantial numerical problem:
one needs to first evaluate the integral numerically,
then substitute this into the generalization of (\ref{ode}) to solve that numerically,
and finally put this into the generalizations of (\ref{et}) to get the pressure and density.

Another problem is what units of length $r$ is measured in.
As galactic rotation curves are observed to be constant at almost all distances
from the centre of galaxies,
the problem is in this sense scale invariant and there is no characteristic
length scale for the problem.
The models here are invariant under $r\rightarrow r/r_0$ so that $r$ is arbitrary,
in particular the models here of galactic rotation do have a characteristic length,
usually when $r=1$ where $\ln(r)$ changes sign,
so the question arises `what is $r=1$ in meters?',
the answer is that this is an arbitrary distance.
An exception is the model (\ref{gvp}) where there is no characteristic length associated
with a sign change of the potential,
but there is the characteristic length of the step at $r_o$.
\section{Conclusion.}\label{conclusion}
That galactic rotation is described by a logarithmic potential,
as opposed to a reciprocal potential $-1/r$,
is sufficient to show that dark matter does not exist.
In particular Fig.2 shows that both the density and mass function
do not tend to zero asymptotically.
Thus galactic rotation requires field modifications as opposed to fluid modifications of
Einstein's equations,  such a field model has been given \cite{mdrst}.
\section{Acknowledgements.}
I would like to thank Jakob Bekenstein,  James Binney,  Antony Fairall,  Alex Feinstein,
Gerry Gilmore,  Tom Kibble,  Andrew Liddle,  David Matravers,
Michael Merrifield and Andrew Taylor for discussion on various aspect of this paper,
and the referee for his comments.

\end{document}